\documentclass[moor,nonblindrev]{informs1}              % for a regular run
\usepackage{natbib}
 \NatBibNumeric
 \def\bibfont{\small}%
 \bibpunct[, ]{[}{]}{,}{n}{}{,}%

\usepackage[colorlinks=true,breaklinks=true,bookmarks=true,urlcolor=blue,
     citecolor=blue,linkcolor=blue,bookmarksopen=false,draft=false]{hyperref}

\def\EMAIL#1{\href{mailto:#1}{#1}}% When hyperref is used, otherwise outcomment 
\def\URL#1{\href{#1}{#1}}         % When hyperref is used, otherwise outcomment 

\TheoremsNumberedThrough     % Preferred (Theorem 1, Lemma 1, Theorem 2)
\EquationsNumberedThrough    % Default: (1), (2), ...
\usepackage{graphicx}
\usepackage{bbm}
\usepackage{booktabs}
\usepackage{tikz}
\usetikzlibrary{matrix}
\usetikzlibrary{arrows,automata}
\usepackage{amsmath,amssymb,url}
\usepackage{algorithm,algpseudocode}
\usepackage{graphics}
\usepackage{float}
\usepackage[T1]{fontenc}
\usepackage[utf8]{inputenc}
\usepackage{subcaption}
\usepackage{tabularx}
\usepackage{verbatim}
\usepackage{pgf}
\usepackage{enumerate}
\usepackage{url}
\usepackage{setspace}
\usepackage{color}
\usepackage{soul}

\def\Pr{\mathbb{P}}

\definecolor{light-gray}{gray}{0.95}
\def\red#1{{\color{red}#1}}

\long\def\old#1{}

\def \cut{\text{c}}
\def \homega{\hat{\omega}}

\def \Pr{\mathbb{P}}

\def \Ex{\mathbb{E}}

\def\cg{c_{\gamma}}
\def\crr{c_{r}}
\def\oM{{\overline{M}}}
\def\hM{{\hat M}}
\def\oB{{\overline{B}}}

\def\barB{\overline{B}}
\def\bt{{\bar t}}

\begin{document}
%%%%%%%%%%%%%%%%

\setcounter{page}{1}
\VOLUME{}%
\NO{}%
\MONTH{}% (month or a similar seasonal id)
\YEAR{}% e.g., 2005
\FIRSTPAGE{}%
\LASTPAGE{}%
\SHORTYEAR{}% shortened year (two-digit)
\ISSUE{} %
\LONGFIRSTPAGE{} %
\DOI{}%

\RUNAUTHOR{Drakopoulos, Ozdaglar, and Tsitsiklis}

 \RUNTITLE{Network resistance}
%\RUNTITLE{}

\TITLE{When is  a network epidemic hard to eliminate?}
%\TITLE{}

\ARTICLEAUTHORS{%
\AUTHOR{Kimon Drakopoulos}
\AFF{LIDS, EECS, MIT, \EMAIL{kimondr@mit.edu}, \URL{http://web.mit.edu/kimondr/www/}}
\AUTHOR{Asuman Ozdaglar}
\AFF{LIDS, EECS, MIT, \EMAIL{asuman@mit.edu}, \URL{https://asu.mit.edu}}
\AUTHOR{John N. Tsitsiklis}
\AFF{LIDS, EECS, MIT, \EMAIL{jnt@mit.edu}, \URL{http://web.mit.edu/jnt/www/home.html}}
} % end of the block

\ABSTRACT{%
We consider the propagation of a contagion process (“epidemic”) on a network and study the problem of dynamically allocating  a fixed curing budget to the nodes of the graph, at each time instant.  For bounded degree graphs, we provide a lower bound on the expected time to extinction under any such dynamic allocation policy, in terms of a combinatorial quantity that we call the resistance of the set of initially infected nodes, the available budget, and the number of nodes $n$.

Specifically, we consider the case of bounded degree graphs, with the resistance growing linearly in $n$. We show that if the curing budget is less than a certain multiple of the resistance, then the expected time to extinction grows exponentially with $n$. As a corollary, if all nodes are initially infected and the CutWidth of the graph grows linearly, while the curing budget is less than a certain multiple of the CutWidth, then the expected time to extinction  grows exponentially in $n$.  The combination of the latter with our prior work  establishes a fairly sharp phase transition on the expected time to extinction (sublinear versus exponential) based on the relation between the CutWidth and the curing budget. }
%

%\KEYWORDS{deterministic inventory theory; infinite linear programming duality; 
%  existence of optimal policies; semi-Markov decision process; cyclic schedule}
%\MSCCLASS{Primary: 90B05; secondary: 90C40, 90C90}
%\ORMSCLASS{Primary: Inventory/production: deterministic multi-item;
%  secondary: dynamic programming/optimal control: deterministic 
%  semi-Markov; programming: infinite dimensional}
%\HISTORY{Received November 20, 2003; revised March 8, 2004, and March 26, 2004.}

\MSCCLASS{}
\ORMSCLASS{Primary: ; secondary: }
\HISTORY{}

\maketitle

% \end{APPENDICES}

\section{Introduction.} \label{intro}
We study the {dynamic} \emph{control} of  contagion processes (from now on called \emph{epidemics}) under limited curing resources. Specifically, we study \emph{dynamic} allocation policies that use information on the underlying {structure} of contacts and on the \emph{infection state} of individuals, and evaluate  performance  in terms of the expected time until the epidemic becomes extinct. In our main contribution, we provide an exponentially large lower bound on the expected  time to extinction, under certain assumptions on the network and the available curing resources.

Our general motivation comes from infectious disease epidemics, although without aiming at a faithful representation of the details of real-world situations. 
One example is the recent outbreak of the Ebola virus which causes an acute and serious illness, which is often fatal if untreated \cite{WHO14}. However, supplies of experimental medicines, e.g.,  the prototype drug ZMapp, are limited and ``will not be sufficient for several months to come,'' {as stated in} \cite{WHO14media}. In view of the limited availability of treatment for the virus, \cite{NYTimes} addresses the following question: ``Ebola Drug Could Save a Few Lives. But Whose?''. Apart from the above, contagion processes are also relevant in the context of {information and influence propagation   
 in social networks \cite{Adad05,Aral12, kim14, Gom10},}
viral marketing \cite{Lesk07}, spread of computer viruses \cite{Gago03}, or diffusion of innovations \cite{Rog03}.

\subsection{Preview of the model.} 
The wide {relevance and} applicability   of contagion processes has led to extensive work on modeling their evolution and on understanding  the resulting dynamics.  Many  models  have been proposed in the literature; see, e.g.,  \cite{Liggett} for an in-depth review of such models and main results. Our work  involves an extension of the canonical SIS epidemic model:  the  epidemic spreads on the underlying network  from an initial set of \emph{infected} nodes to \emph{healthy} nodes {and at the same time, infected nodes can be cured.}  
Healthy nodes get infected at a constant and common \emph{infection rate} by each of their infected neighbors. In contrast to the standard SIS model, which assumes  a common curing rate for all infected  nodes at all times, we assume instead   a node   and time-specific \emph{curing rate}.   A \emph{curing policy}, to be applied by a central controller, is a choice, at each time instant, of the curing rates at each node, taking into account the history of the epidemic and the network structure, subject to a budget on the sum of the curing rates applied at each time. 
The resulting  process  is  a  controlled finite Markov chain with a unique absorbing state:  the state where all nodes are healthy. We say that the epidemic becomes \emph{extinct} when that absorbing state is reached. Under mild assumptions on the  curing budget and given any set of initially infected nodes, the epidemic becomes extinct in a random but finite amount of time. The main question here is how much time will be needed.

We can draw  an important qualitative distinction between networks in which (i) the spread of the epidemic is hard to stop with the given curing budget, so that the expected time to extinction  grows exponentially with the number of nodes, and (ii) networks for which the curing resources are adequate, so that the expected time to extinction  grows slowly (polynomially or even sub-polynomially) with the number of nodes. Our general objective is to develop criteria that allow us to distinguish between cases (i) (slow extinction) and (ii) (fast extinction). In this paper,
we focus on graphs in which the maximum degree is bounded by some $\Delta$ (independent of $n$) and provide the answer for a particular ``regime'', {namely, the case} {of graphs whose  CutWidth grows linearly}.

\subsection{Main contribution.}\label{s:contrib} In a companion paper \cite{DOT14}, we have established that a certain combinatorial quantity, the CutWidth of the underlying graph, denoted by $W$, plays a central role. In particular, if the curing budget $r$ satisfies $r\geq 4W$ and $r\geq 16\Delta\log_2n$, where $n$ is the number of nodes, then the expected time to extinction  is small --- in fact, upper bounded by $26n/r$, and therefore sublinear in $n$. In the present paper, we establish  {a} converse {result}, for graphs with large CutWidth, {namely,} for graphs whose CutWidth is lower bounded by $\cg n$, for some constant $\cg>0$. In particular, we show that if $r\leq \crr W$, where $\crr>0$ is an absolute constant (depending only on the degree bound and on $\cg$), then, for some initial states, the expected time to extinction  is at least exponential, under any curing policy. In other words, for graphs whose CutWidth scales linearly with $n$, a curing budget that also scales linearly with $n$ is necessary (from the results in this paper) and sufficient (from the results in \cite{DOT14}) for fast extinction.
In an equivalent interpretation of our main result, we are establishing that, {for the case of bounded degree graphs,} fast extinction with a sublinear curing budget is possible if and only if the CutWidth grows sublinearly.

\subsection{Related literature.}
A similar {problem, but in which the curing rate allocation is  static (open-loop)} has been studied in \cite{Cohen,gourdin, chung,preciado}, 
but the proposed methods were either heuristic or based on  mean-field approximations of the evolution process; see \cite{preciado2} for a survey.  {Closer} to our work, {the authors of  \cite{Borg10} let the curing rates be proportional} to the degree of each node and independent of the current state of the {network, which may actually result in  having curing resources wasted on healthy nodes.} For  bounded degree graphs, the policy in \cite{Borg10} achieves sublinear expected time to extinction, but requires a curing budget that is proportional to the number of nodes. In contrast, the dynamic policy in \cite{DOT14} achieves the same performance (sublinear expected time to extinction) for  bounded degree graphs with small CutWidth, \emph{more economically},  by properly allocating a sublinear curing budget, hence demonstrating the increased effectiveness of dynamic policies.

Regarding lower bounds for dynamic policies, \cite{Borg10} establishes that for expander graphs, and with a sublinear curing budget, the expected  time to extinction is at least exponential in the number of nodes, under any curing policy. 
Expander graphs have automatically a large CutWidth, and so
our result is in the same flavor, but much more general and also much harder to establish. The argument behind the result in \cite{Borg10} is essentially the following:  for expander graphs, any sufficiently large set of infected nodes results in a large (linear) total infection rate, which cannot be countered by a sublinear curing budget. But more general graphs with a large CutWidth need not have such a property: it may be the case that the instantaneous total  infection rate is large for only ``a few'' configurations (sets of infected nodes). For such graphs, one can still establish that the total infection rate will be larger than the curing rate at some point in time, but this may last, in principle, for only a small time interval, and this is not enough to establish a negative result, in the form of a strong lower bound. We {finally} mention \cite{WaAn05}, which also deals with dynamic policies, but for the special case of a line graph.

In order to establish our result, we have to argue that a large total infection rate (larger than the curing budget) will be encountered for a sufficiently long time interval, and that this creates a barrier to the fast extinction of the epidemic. The argument involves an elaborate combinatorial analysis of the evolution of the set of infected nodes. 

We finally note a related negative result (exponential expected time to extinction) that we have established in \cite{cdc15}. That result deals with a special case, namely, graphs for which the CutWidth is close to the largest possible value for graphs of the given size. The result in \cite{cdc15} admits however a much simpler proof because under the large CutWidth assumption, it is easier to identify a barrier (a situation where the instantaneous total  infection rate is high) inside which the process must remain for a sufficiently long time.

\subsection{Outline of the paper.}
The rest of the paper is organized as follows. In Section \ref{model} we present the epidemic propagation model and define the curing policies under consideration. In Section \ref{sec:combinatorics} we define the 
CutWidth, as well as a generalization of that concept, and develop some combinatorial preliminaries that will be needed later.
Section \ref{s:result} contains a statement of the main result, two key lemmas that comprise the core of the proof, and some discussion.
Sections \ref{s:l1}-\ref{s:l2} contain the proofs of the two key lemmas.
Finally, Section \ref{s:con} contains some concluding remarks.

\section{The Model.} \label{model}
The model that we use and the contents of this section are borrowed from \cite{DOT14} and \cite{cdc15}. We consider a {network}, represented by an {undirected} graph  $G=(V,E)$, where $V$ denotes the set of nodes and  $E$ denotes the set of edges. {We use $n$ to denote the number of nodes.} Two nodes $u,v \in V$ are {\it neighbors} if $(u,v) \in E$. 
We restrict to graphs for which the node degrees are {upper bounded} by $\Delta$, which we take to be a given constant throughout the paper.

We  let $I_0{\subseteq V}$ be a set of intially infected nodes, and assume that the infection spreads according to a controlled contact {(or SIS)} process, where the rate at which infected {nodes} get cured is  determined by a network controller. Specifically, each {node} can be in one of two states: {\it infected}  or {\it healthy}. The controlled contact process 
 is a 
{right-continuous}, continuous-time Markov process 
{$\{I_t\}_{t\geq 0}$ on the state space $\{0,1\}^V$, where $I_t$ stands for the set of infected nodes at time~$t$.} {We refer to $I_t$ as the {\it infection process}.} We will sometimes use  $I_{t-}$ as a short-hand for the value $\lim_{s\uparrow t}I_s$ just before time $t$.

At any point in time, state transitions at each node occur independently, according to the following {rates. (These rates essentially define the generator matrix of the continuous-time Markov process under consideration.)} 
\begin{enumerate}[a)] \item {The process 
is initialized at the given initial state $I_0$.}
\item If a node $v$ is healthy, i.e., if $v\notin I_t$, the transition rate associated with a change of the state of that node to being infected is  equal to a positive infection rate $\beta$ times the number of infected neighbors of $v$, that is,
$$\beta \cdot  \big|\{(u,v)\in E: u\in I_t\}\big|,$$
where we use $|\cdot|$ to denote the cardinality of a set.
Any transition of this type will be referred to as an {\it infection}.
By rescaling time, we can and will assume throughout the paper that $\beta=1$. 

\item {If a node $v$ is infected, i.e., if $v\in I_t$, the transition rate associated with a change of the state of that node to being healthy is equal to a curing} rate  
 $\rho_v(t)$ that is determined by {the} network controller, as a function of  the current and past states of the {process.} We are assuming here that the network controller has access to {the entire} history of the process.
Any transition of this type will be referred to as a \it{recovery}.
\end{enumerate}

We impose a {\it budget constraint} 
{of the form}    
\begin{equation}
\sum_{v \in V} \rho_v(t)\leq r,  \label{eq:budgetconstr}
\end{equation}
{for each} time instant $t$, reflecting the fact that curing is costly. 
 A {\it curing policy}  is a  mapping {which at any time $t$ maps the past history of the process to a curing vector $\rho(t)=\{\rho_v(t)\}_{v \in V}$} that satisfies~(\ref{eq:budgetconstr}). 
 
We define the {\it  time to extinction} as the first time when the process first reaches the absorbing state where all nodes are healthy: 
$$
\tau = {\min}\{t\geq0 : {I_t=\emptyset}\}. 
$$			
{In this paper,} {we focus on} the  {\it expected  time to extinction} (the expected value of $\tau$),  {as} the performance measure {of interest.}

Without loss of generality,  we can and will restrict to curing policies that allocate the entire budget $r$ to infected nodes, as long as such nodes exist; this is because  {having unused curing resources or allocating them to healthy nodes would be wasteful.}
  Under this restriction, 
the empty set (all nodes being healthy) is a unique absorbing state, and therefore the  time to extinction is finite, with probability~1.

Finally, we can and will restrict to policies that at any point in time allocate the entire budget to a single infected node, if one exists. We can do this because it is not hard to show that there exist optimal policies (i.e., policies that minimize the expected time to extinction) with this property.\footnote{{A formal proof of this statement (which we only outline) goes as follows. We write down the Bellman equation for the problem of minimizing the expected time to extinction and observe that the right-hand side of Bellman's equation is linear in 
$\rho(t)$. We then recall that $\rho(t)$ is constrained to lie in a certain simplex, and conclude that we can restrict, without loss of optimality, to the vertices of that simplex. Any such vertex corresponds to allocating the entire budget to a single infected node.}}

\section{Graph theoretic preliminaries.}\label{sec:combinatorics}
In this section, after some elementary definitions and notation, we focus on a deterministic version of the problem under consideration. Variants of such  deterministic problems have been studied in the literature \cite{LaPaugh93,Pars78} and  {involve} the concept of the \emph{CutWidth} of a graph. Loosely speaking, the CutWidth is the maximum cut encountered during the \emph{deterministic} extinction of an epidemic on a graph, starting from all nodes infected, in the absence of any {reinfections of nodes that have become healthy}, and under the best possible sequence with which nodes are cured. (A formal definition will be given shortly.)

We also introduce and study a natural extension of the concept of the CutWidth, for the case where only a subset of the nodes is initially infected;  we refer to it as the \emph{resistance} of the subset.The resistance turns out to  contain  important information about  the evolution {of} an epidemic,  starting from the corresponding subset, and will  serve as a  low-dimensional summary of the state of an infection process. 
In the  {sub}sections that follow, we introduce those two concepts and study the properties of the latter.   

\subsection{Notation and Terminology.}
For convenience, we use the term \textbf{bag} to refer to {a} ``subset of $V$.'' For any bags $A$, $B$, and any node $v$, we define
$$
A \setminus B= \{v \in A: v \notin B\},
$$
{which is} {the set of nodes that belong in $A$ but not in $B$, and}
$$
\qquad A\triangle B = (A \setminus B) \cup (B \setminus A),
$$
{which is} {the set of nodes at which $A$ and $B$ differ.}
Finally, we write
\[
A+v=A\cup\{v\}, \qquad
A-v=A \setminus \{v\}.
\]

We next define the concept of a crusade from $A$ to $B$ as a sequence of bags that starts \red{at} $A$ and ends at $B$, with the restriction that at each step of this sequence, arbitrarily many nodes may be added to the previous bag, but at most one can be removed. The formal definition follows.

\begin{definition}
For any two bags $A$ and $B$, an  $(A$-$B)$-\textbf{crusade} $\omega$  is a sequence $(\omega_0,\omega_1,\ldots,\omega_{k})$ of bags, of length $k+1$, with the following properties:
\begin{enumerate}[(i)]
\item $\omega_0 = A$,
\item $\omega_{k}=B$, and
\item $|\omega_i \setminus \omega_{i+1} |\leq 1,$ for  $i ={0}, \ldots, k-1.$
\end{enumerate}

{We use the notation $\Omega(A)$ to refer to the set of all 
($A$-$\emptyset)$-crusades, i.e., crusades that start with a bag $A$ and eventually end up with the empty set.}
\end{definition}

Property (iii) states that at each step of a crusade, arbitrarily many nodes can be added to, but \emph{at most one} node can be removed from the current bag.
Note that the definition of a crusade allows for \emph{non-monotone} changes, since a bag at any step can be a subset, a superset, or  not comparable to the preceding bag. For this reason, crusades, as defined here, are different from the  \emph{monotone} crusades that were introduced in \cite{DOT14}. 

\subsection{Cuts, CutWidth, and Resistance.}
The number of edges connecting {a} bag {$A$} with its complement will be called the cut of the bag. {Its importance lies in that it is equal to the total rate at which new infections occur, when the set of currently infected nodes is $A$.}
\begin{definition} {For any bag $A$, its \textbf{cut}, $\cut(A)$, is defined as the cardinality of the set of edges
$$\big\{(u,v): u\in A,\ v\in A^c\big\}.$$}
\end{definition}

\noindent 
In {Lemma} \ref{prop:cutproperties} below, we record, without proof, some  elementary properties of cuts.

\begin{lemma}
\label{prop:cutproperties}
For any two bags $A$ and $B$, we have 
\begin{itemize}
\item[(i)] {$\big|\cut(A)-\cut(B)\big| \leq \Delta \cdot\big| A\triangle B\big|$.}
\item [(ii)] If $A\subseteq B$, and  $v \in A $, then
\[
\cut(A-v)  - \cut(A) \leq \cut(B-v) - \cut(B).
\]
%\item[(iii)]	 $\cut(A)\leq \min\{|A|\cdot \Delta, (n-|A|)\cdot \Delta\}$.
%\red{[DO WE NEED THIS?]}
\end{itemize}
\end{lemma}
Note that Lemma \ref{prop:cutproperties}(ii) states the well-known submodularity property of the function $\cut(\cdot)$, and thus of the infection rate.

We  now define  the width of a crusade  as the maximum cut that it encounters. 
\begin{definition}
The \textbf{width} $z(\omega)$ of  an ($A$-$B$)-crusade $\omega=(\omega_0, \ldots, \omega_{{k}})$ is defined by
\[
z(\omega)=\max_{ 1 \leq i \leq {k}}\{\cut(\omega_i)\}.
\]
\end{definition}
Note that in the above definition, the maximization starts at the first step of the crusade, i.e., we exclude $\omega_0$ from consideration. The reason is the important Monotonicity property in Lemma \ref{lem:infectiondelta}(i), in the next subsection, which would otherwise fail to hold.

{We finally define the \emph{resistance}\footnote{Note that in \cite{DOT14} we used a related notion, where the  minimization took place with respect to monotone crusades.}  of a a bag $A$ as the minimum crusade width, over all ($A$-$\emptyset$)-crusades.  Intuitively, this is the  maximum cut  encountered after the first step, during a crusade that ``cures'' all nodes in $A$ in an ``optimal'' manner. }

\begin{definition}  The \textbf{resistance} $\gamma(A)$ of a bag $A$ is defined  by  
$$
\gamma(A)= \min_{\omega \in \Omega(A)} z(\omega). 
$$
{For the special case where $A$ is the set $V$ of all nodes, the corresponding resistance $\gamma(V)$ is called the \textbf{CutWidth} of the graph and is denoted by $W$.}
%\[
%\Omega^A=\{ \omega \in \Omega(A): z(\omega)=\gamma(A) \}.
%\]
 %Crusades in $\Omega^A$ are referred to as $A$-optimal. 

\end{definition}

  We remark that the more common definition of the CutWidth, {and} which was the one used in 
\cite{DOT14}, takes the minimum over monotone crusades, i.e., over crusades that only remove nodes. 
Nevertheless, the two definitions are equivalent, as has been established in
 \cite{BiSe91} and \cite{LaPaugh93}.

We {close this section by observing}  that the resistance  of a bag $A$ satisfies the Bellman equation 
\begin{equation} \label{eq:bellman}
\gamma(A) = \min_{|A\setminus B|\leq 1} \big\{ \max\{\cut(B), \gamma(B)\}\big\}.
\end{equation}

\subsection{Properties of the resistance.}\label{sec:resil}
This section develops some properties of the resistance.  Lemma \ref{lem:infectiondelta}(i) states that if $A$ and $B$ are two bags with  $A \subseteq B$, then $\gamma(A) \leq \gamma(B)$. Intuitively, this is because one can construct a crusade from $A$ to $\emptyset$ as follows:  The crusade starts from  $A$, then continues to the first bag encountered by a $B$-optimal crusade $\omega^B$,  and then follows $\omega^B$. The constructed crusade and $\omega^B$ are the same except for the respective initial bags. By the definition of the resistance, the initial bag does not affect the maximization  and thus the {width of the} new crusade  is equal to $\gamma(B)$.  An optimal crusade from $A$ can do no worse.

Lemma \ref{lem:infectiondelta}(ii) states that if two bags $A$ and $B$ differ by only $m$ nodes, then the corresponding resistances are at most $m\Delta$ apart. Intuitively, this is because if $m=1$ and $A\triangle B=\{v\}$, one can attach node $v$  to the optimal crusade {for} the {smaller} of the two bags, thus obtaining a crusade that starts at the larger bag and encounters a maximum cut which is at most $\Delta$ different from the original. The result for general $m$ is obtained by moving from $A$ to $B$ by adding or removing one node at a time. 

The formal proof of Lemma \ref{lem:infectiondelta} follows the above outlined intuitive argument, and is given in Appendix
\ref{ap:lemma}, so { as not to} disrupt continuity.
%\begin{figure}
%\centering
%\includegraphics[width=0.6\textwidth]{monotonicityproof}
%\caption{Proof of the monotonicity of the resistance. Solid lines are used for optimal crusades while dotted for the constructed crusade.} \label{fig:monotonicityproof}
%\end{figure}

\begin{lemma}\label{lem:infectiondelta}
{Let $A$ and $B$ be two bags.}
\begin{enumerate}[(i)]
\item {[Monotonicity]} {If} $A \subseteq B$, then  $\gamma(A) \leq \gamma(B)$.
\item {[{Smoothness}]}
{ We have that $\big|\gamma(A)-\gamma(B)\big| \leq \Delta \cdot\big| A\triangle B\big|$.}
%{If} $B=A+v$, then $\gamma(B) \leq \gamma(A)+\Delta$.
\end{enumerate}
\end{lemma}

{An immediate corollary of Lemma \ref{lem:infectiondelta}(i) is that for any bag $A$, we have $\gamma(A)\leq W$.}

\paragraph{Example.} Consider a line graph with $n$ nodes. Its CutWidth is easily seen to be equal to 1: {if all nodes are initially infected,} we can cure {them} one at a time, starting from the left;   the cuts encountered {along} the way are all equal to 1. On the other hand, if all even nodes are initially infected, the corresponding cut is large, equal to $n-1$. However, the cut being large does not acurately convey the difficulty of curing those nodes. For example, we might artificially infect the healthy nodes (or, in the stochastic SIS model, simply wait until they all get infected---{this would happen {in time which is sublinear in $n$),} and then follow {a} curing policy for the case of a fully infected initial graph. Thus, the expected time to extinction {will be} {comparable to} {the one for} the case where all nodes are initially infected. {Note} that the resistance  {of any nonempty bag} is equal to 1: {an} optimal {non}monotone crusade can start by infecting all nodes {(except possibly one of the end nodes, if $n$ is even), and then curing the nodes one at a time.} 
{Thus, the resistance, rather than the cut, is a better reflection of the difficulty of curing a given initial set of infected nodes.}	
	As a side-note, these considerations suggest
that some additional infections in the beginning (as allowed in our definition of the resistance) can be beneficial, {and this is one of the reasons why our line of argument is based on nonmonotone (as opposed to monotone) crusades.}

\subsection{Relating cuts to the resistance.}

{As illustrated in the preceding example, a large value of $\cut(I_0)$ does not mean that the infection is hard to extinguish; the resistance is more relevant. On the other hand, the proof of any negative result (i.e., a lower bound on the expected time to extinction) has to argue that at certain times the cut will be large and will present a barrier to the extinction of the epidemic. For this reason, 
we need a way of pinpointing certain times at which  a large resistance implies a large cut. This is accomplished by the next lemma, which establishes a connection between cuts and resistances at those times that the resistance is reduced.
It shows that whenever the resistance is high and is reduced, the total infection rate is  also high. This observation will play a central role in the proof of our main result.}

%\begin{definition} \label{def:improvement}
%Let $A$ be a bag and suppose that $\gamma(A-v)<\gamma(A)$, for some $v\in A$. We then say that $A$ is an {\textbf{improvement}} bag.
%\end{definition}

%Improvement bags have the important property that their cut can be approximately lower bounded by their resistance. 

% \begin{lemma} \label{lem:cutwhendrops}
%For any  improvement bag, we have that 
%\[
%\cut(A) \geq \gamma(A)  -\Delta.
%\]
 %\end{lemma} 
 
 \begin{lemma} \label{lem:cutwhendrops}
 Let $A$ be a bag and suppose that $\gamma(A-v)<\gamma(A)$, for some $v\in A$. Then,
 $${\cut(A-v) \geq \gamma(A).}$$
 \end{lemma} 
\proof{Proof:}
Let $B=A-v$.
Since $|A\setminus B|=1$,  Eq.~\eqref{eq:bellman} implies that
\begin{equation}
\gamma(A) \leq \max \{\cut(B), \gamma(B)\}. \label{eq:drop}
\end{equation}
Having assumed that $\gamma(B)<\gamma(A)$,  Eq.~(\ref{eq:drop}) implies  that $\gamma(A) \leq \cut(B)$.  \Halmos
\endproof

\section{The main result and the core of its proof.}\label{s:result}
In this section we state our main result and provide the key elements of its proof in the form of two  lemmas. Loosely speaking, the result states that if the resistance of the initial bag scales linearly with the number $n$ of nodes, and the budget scales only as a small constant multiple of $n$, then the expected time to extinction is exponentially large.

\begin{theorem}\label{th:main}
Consider a graph with $n$ nodes and a set $I_0$ of initially infected nodes, 
and suppose that {for some constant $c_\gamma$},
$$\gamma(I_0)\geq \cg n.$$
Suppose, furthermore that all node degrees are bounded above by  $\Delta$. 
Then, there exist positive constants $
\crr$ and $c$, which only depend on $\cg$ and $\Delta$, such that if
$$r\leq \crr n,$$
then
$$\Pr\Big(\tau \geq ce^{cn}\Big) \geq \frac{1}{2},$$
under any policy, and for all large enough $n$. In particular,
$$\Ex[\tau]\geq \frac{1}{2} ce^{cn}.$$
\end{theorem}

\noindent
{{\bf Remark:} An immediate corollary of Theorem \ref{th:main} is obtained by letting $I_0=V$, so that $\gamma(I_0)$ coincides with the CutWidth $W$:  if the CutWidth scales linearly in $n$, and the curing budget is less than a certain multiple of the CutWidth, then the expected time to extinction grows exponentially in $n$. As a further corollary, if the curing budget grows sublinearly with $n$, fast extinction is possible only if the CutWidth  grows sublinearly in $n$. This is a converse to the results of \cite{DOT14}, which establish that if the CutWidth  grows sublinearly in $n$, then fast extinction is possible with a sublinear budget.
}

 The proof of our main result involves the following line of argument.
\begin{itemize}
\item[a)] In the first, deterministic, part of the proof (Lemma \ref{biglemma1}), we show that  {for graphs with large CutWidth, 
the time interval until the extinction of the epidemic
must contain a substantially long  subinterval} during which the expected total infection rate is significantly larger than the budget, yet the realized ratio of infections to recoveries is relatively small, and in particular, fairly different than the ratio of the corresponding expected rates.
\item[b)] In the second, stochastic, part of the proof (Lemma \ref{biglemma2}), 
we argue that for a given time interval to have the properties in a), a ``large deviations'' event, with exponentially small (in $n$) probability, must occur. This is used to conclude that, with significant probability, it will take an exponentially long amount of time until an interval with the properties in a) emerges.
\end{itemize}

%\begin{itemize}
%\item[a)] In the first, deterministic, part of the proof (Lemma \ref{biglemma1}), we show that  any sample path \sout{that  extinguishes the epidemic} must contain a substantially long time interval during which the expected total infection rate is significantly larger than the budget, yet the realized ratio of infections to recoveries is relatively small, and in particular, fairly different than the ratio of the corresponding expected rates.
%\item[b)] In the second, stochastic, part of the proof (Lemma \ref{biglemma2}), 
%we argue that for a given time interval to have the properties in a), a ``large {deviations}'' event, with exponentially small (in $n$) probability, must occur. This is used to conclude that, with significant probability, it will take an exponentially long amount of time until an interval with the properties in a) emerges.
%\end{itemize}

\noindent
{\bf Proof of Theorem \ref{th:main}.} 
We start the proof by fixing a graph with $n$ nodes, and the initial set $I_0$ of infected nodes. For convenience, from now on, we will use the short-hand notation  $\gamma$ instead of $\gamma(I_0)$. We assume that $\cg$ and $\Delta$ have been fixed, and that  $\gamma\geq \cg n$. 
Note that for sufficiently large $n$, $\gamma$ will be much larger than $\Delta$, so that we can use freely inequalities such as $\Delta<\gamma/4$, or $\gamma/4\Delta>1$.
In order to keep notation simple and avoid the use of ceilings and floors, we will also assume from now on that $\gamma/4\Delta$ is an integer.  The proof for the general case, is essentially the same.

The first part of the proof corresponds to the following lemma.
\begin{lemma}\label{biglemma1}
Consider a sample path for which $\tau<\infty$. For that sample path, there exist times $t'$ and {$t''$}, with $0\leq t'\leq t'' \leq \tau$, such that:
\begin{itemize}
\item[(i)] $\cut(I_t)\geq \gamma/4$, for all $t\in [t',t'']$; 
\item[(ii)] we have  $b=(\gamma/4\Delta)-1$ recoveries during the interval $[t',t'']$;
\item[(iii)] we have no more than $n+b$ infections during the interval $[t',t'']$.
\end{itemize}
\end{lemma}

The times $t'$ and $t''$ in the preceding lemma are random variables (they depend on the sample path). However, they are not necessarily stopping times of the underlying stochastic process.

Note that it suffices to prove the existence of a time interval $[t',t'']$ with just properties (i) and (ii). This is because there are only $n$ nodes in the graph. If we have $b$ recoveries during a time interval, the number of infections cannot exceed $n+b$, and property (iii) follows automatically.

For the stochastic part of the proof, let us introduce some notation: for any $c>0$, we define $B_c$ to be the event that there exist times $t'$, $t''$, with the properties in Lemma \ref{biglemma1}, together with the additional property $t''\leq ce^{cn}$.

\begin{lemma}\label{biglemma2}
Having fixed $\cg$ and $\Delta$, there exist small enough positive constants $\crr$ and $c$ such that if
$r\leq \crr n$, then
$$\Pr(B_c) \leq\frac{1}{2},$$
for all large enough $n$.
\end{lemma}

Lemmas \ref{biglemma1} and \ref{biglemma2} immediately imply Theorem 
\ref{th:main}. To see this, Lemma \ref{biglemma1} implies that $t''$ is well defined for any sample path. For any sample path that satisfies $\tau\leq ce^{cn}$, we must also have $t''\leq ce^{cn}$. Thus, the event
$\{\tau\leq c e^{cn}\}$ is a subset of the event $B_c$. Using Lemma
\ref{biglemma2}, we conclude that $\Pr(\tau\leq c e^{cn}) \leq \Pr(B_c) 
\leq 1/2$, as long as $\crr$ and $c$ are suitably chosen.

\section{Proof of Lemma \ref{biglemma1}.}\label{s:l1}

Lemma 4 is the central — and least obvious — part of the proof. Before continuing with
a formal argument, we provide a high-level informal overview, intended to enhance comprehension.
The overall plan is to argue that $\gamma(I_t)$, whose initial value is $\gamma$, must eventually (at some time $T$) drop to $\gamma/2$, and that while the value $\gamma/2$ is approached, there must be a sufficiently long interval during which $\cut(I_t)$ is  at least $\gamma/4$. Indeed, if $\cut(I_t)\geq \gamma/4$ for all times in $[0,T]$ (this is Case 1 below), the cut remains relatively large (and larger than the budget), which implies that the process is moving in a direction opposite to its drift; in particular, the probability of this happening is small.

Recall now that the cut is approximately equal to the resistance at those times that the resistance drops. Thus, $\cut(I_T)$ is approximately equal to $\gamma/2$.
If $\cut(I_t)$ drops below $\gamma/4$ before time $T$ (this is Case 2 below), there must exist an interval $[T’,T]$ during which $\cut(I_t)\geq \gamma/4$, and during which the cut increases from $\gamma/4$ to $\gamma/2$. We want to argue that such an increase must be accompanied by a large number of recoveries (which will consist a low-probability event). The difficulty is that cut increases may be caused by either recoveries or infections. In order to isolate the effects of recoveries, we look at a ``bottleneck process’’ $\Theta_t$ that starts the same as $I_t$ at time $T’$, and which keeps track of the recoveries in $I_t$, while ignoring the infections. Similar to $I_t$, there will be a time at which the resistance of $\Theta_t$ will drop to $\gamma/2$ (this is due to the fact that $\Theta_t\subset I_t$, and monotonicity), and at that time, $\cut(\Theta_t)$ will be roughly equal to $\gamma/2$. Thus, $\cut(\Theta_t)$ also increases from $\gamma/4$ to $\gamma/2$. However, because $\Theta_t$ only changes whenever the process $I_t$ has a recovery, it follows that there must be $O(\gamma)$ recoveries in the process $\Theta_t$ and, therefore, for the process $I_t$ as well  (Lemma \ref{l:crux}).

We can now start with the formal proof.
Let us fix a particular sample path for which $\tau<\infty$. 
Let $T$ be the first time that $\gamma(I_t)$ drops to a value of $\gamma/2$ or less:
$$T=\inf\big\{t\geq 0: \gamma(I_t)\leq \gamma/2\big\}.$$
Given that $\gamma(I_{\tau})=\gamma(\emptyset)=0$, it is clear that such a time $T$ exists and satisfies $0\leq T\leq \tau$.

We distinguish between two cases:
\par\noindent
{\bf Case 1:}
Suppose that throughout the interval $[0, T]$, we also have $\cut(I_t)\geq \gamma/4$.
Because of the monotonicity property of $\gamma(\cdot)$ (Lemma \ref{lem:infectiondelta}(i)), $\gamma(I_t)$ decreases only when the set $I_t$ decreases, that is, only when there is a recovery. Furthermore, using the smoothness property in Lemma \ref{lem:infectiondelta}(ii), each time that
there is a recovery, $\gamma(I_t)$ can drop by at most $\Delta$. Therefore,   the number of recoveries during the time interval $[0,T]$  is at least 
$$\frac{\gamma(I_0)-\gamma(I_T)}{\Delta}
\geq
\frac{\gamma-\gamma/2}{\Delta}=\frac{\gamma}{2\Delta}.$$ 
We can then find some $\hat T\leq T$ such that during the time interval $[0,\hat T]$, we {have} exactly $\gamma/4\Delta-1$ recoveries, and
properties (i)-(ii)  in the statement of Lemma \ref{biglemma1} are satisfied by letting $t'=0$ and $t''=\hat T$. 

\par\noindent
{\bf Case 2:}
Suppose now that there exists some $t\in[0, T]$, with $\cut(I_t)< \gamma/4$, which is the more difficult case.
Note that just before time $T$, we have $\gamma(I_{T-})>\gamma/2$. Furthermore, $\gamma(I_T)\leq \gamma/2$. 
With our continuous-time Markov chain model, only one event (infection or recovery) can happen at any  time. Since $\gamma(I_T)<\gamma(I_{T-})$, and since $\gamma(\cdot)$ is monotonic, it follows that we had a recovery and, therefore, $I_T=I_{T-}-v$, for some node $v$.
Lemma \ref{lem:cutwhendrops} applies, with $A=I_{T-}$ and $A-v=I_T$, and we obtain
$$\cut(I_T)\geq \gamma(I_{T-})>\frac{\gamma}{2}.$$
We now define
$$T'=\sup\{t\leq T: \cut(I_t)<\gamma/4\},$$
so that $\cut(I_{T'-})<\gamma/4$. 
%\red{ \st{and therefore $T'\leq T$} THIS IS IN THE DEFINITION}. 
Furthermore, 
since $\cut(I_t)$ can change by at most $\Delta$ at each transition (Lemma 
\ref{prop:cutproperties}), we must actually have 
\begin{equation}\label{eq:cutt}
\cut(I_{T'})<\frac{\gamma}{4}+\Delta,
\end{equation}
 which implies that $T'\neq T$ and $0\leq T'< T$.

We {will show} that the interval $[t',t'']$, with $t'=T'$ and $t''=T$ has 
properties (i)-(ii)  in the statement of Lemma \ref{biglemma1}. Indeed, 
the definition of $T'$ implies that
$$\cut(I_t)\geq \frac{\gamma}{4},\qquad \forall\ t\in[T',T],$$
which is property (i). 
The proof of Lemma \ref{biglemma1} is completed by showing property (ii), namely, that
the increase in $\cut(I_t)$, from a value smaller than $(\gamma/4)+\Delta$ (at time $T'$), to a value above $\gamma/2$ (at time $T$) together with a drop of the resistance from a value above $\gamma/2$ (at time $T'$) to a value below $\gamma /2$ (at time $T$), must be accompanied by at least $(\gamma/4\Delta)-1$ recoveries.
This is the content of the next lemma.

\begin{lemma}\label{l:crux}
The number of recoveries during the time interval $[T',T]$  is at least
$$
\frac{\gamma}{4\Delta} -1.$$
\end{lemma}

Lemma \ref{l:crux} is a rather simple statement, but we are not aware of a simple proof {or of a transparent intuitive explanation.} 
 Our proof relies on an auxiliary process, the \emph{bottleneck process}, coupled with $I_t$, which is introduced and analyzed in the next section.

\section{Proof of Lemma \ref{l:crux}.}\label{sec:bottleneck}

{The first step in proving Lemma \ref{l:crux} is the construction of}  a process which is coupled with the infection process. Observe that a sample path of the infection process defines a crusade in which, at each step, a single node is added to or removed from the current bag. To any such crusade, we associate  a \emph{bottleneck sequence}, which is  a sequence of bags consisting of subsets of the bags in the  original crusade, with several important properties. 
Consider a crusade $\omega=(A_0,A_1, \ldots, A_l)$ in which  $|A_{i} \triangle A_{i-1}|=1$, for $i =1 ,\ldots, l$.  
{In particular, we always have $A_i\subset A_{i-1}$ or $A_i\supset A_{i-1}$.} 
We associate with  $\omega$ a related sequence of bags $(\Theta_0, \dots, \Theta_l)$, 
{by letting}
\begin{equation}
\Theta_i= \bigcap_{k=0}^{i} A_k,\qquad
{i=0,\ldots,l.} 	\label{eq:bot}
\end{equation}

It is clear from  our construction  that $\Theta_i$ is always a subset of $A_i$, and that $\Theta_i\supseteq \Theta_{i-1}$. We  have the following interpretation: $\Theta_0$ starts the same as $A_0$. Whenever a node is removed from a bag in the original sequence,  the same is done in the bottleneck sequence, as long as this is possible. On the other hand, whenever a node is added to a bag in the original sequence, nothing is done in the bottleneck sequence. 
\begin{lemma} \label{prop:bottleneck}
Consider a sequence  $(A_0,A_1, \ldots, A_l)$  of bags such that  $|A_{i} \triangle A_{i-1}|=1$, for $i=1,\ldots,l$, and the associated bottleneck sequence $(\Theta_0,\ldots,\Theta_l)$. The following hold:
\begin{enumerate}[(i)]
\item $\Theta_i \subseteq A_i$.
\item If $\cut(\Theta_{i}) > \cut(\Theta_{i-1})$, then $A_{i} \subset A_{i-1}$. 
\item $\cut(\Theta_{i}) - \cut(\Theta_{i-1}) \leq \Delta$.
\end{enumerate}
\end{lemma}

\proof{Proof:}
\begin{enumerate}[(i)]
\item Follows directly from the definition.

\item 
{Suppose that $\cut(\Theta_{i}) > \cut(\Theta_{i-1})$. Then, $\Theta_i\neq \Theta_{i-1}$. From the definition of the bottleneck sequence, we see that it if $A_i\supset A_{i-1}$, then $\Theta_i=\Theta_{i-1}$. Therefore, we must have that $A_i\subset A_{i-1}$.} 

\item If $A_{i} {\supset} A_{i-1}$, then,  $\Theta_i=\Theta_{i-1}$,  and $\cut(\Theta_{i}) - \cut(\Theta_{i-1})=0$. On the other hand, if $A_i \subset A_{i-1}$, and using the assumption $|A_{i} \triangle A_{i-1}|=1$, we write $A_i=A_{i-1} - v$ for some $v \in A_{i-1}$, and from Eq. (\ref{eq:bot}) we obtain $\Theta_{i}=\Theta_{i-1}-v$. The result then follows from Lemma \ref{prop:cutproperties}{(i)}. \hfill \Halmos

\end{enumerate}

\endproof

We now complete the proof of Lemma \ref{l:crux}.
Let $A_0,\ldots,A_l$ be the sequence of bags that arise during the evolution of $I_t$, between times $T'$ and $T$. In particular, $A_0=I_{T'}$
and $A_l=I_T$. Let $\Theta_0,\ldots,\Theta_l$ be the corresponding bottleneck sequence, so that $\Theta_0=A_0=I_{T'}$. Using property (i) in Lemma 
\ref{prop:bottleneck}, we have $\Theta_i\subseteq A_i$, for all $i$. Using the nonotonicity of $\gamma(\cdot)$, we obtain $\gamma(\Theta_i)\leq \gamma(A_i)$, for all $i$. In particular, 
$$\gamma(\Theta_l)\leq \gamma(A_l)=\gamma(I_T)\leq\frac{\gamma}{2}
< \gamma(I_{T'})=\gamma(\Theta_0).$$
(The second and third inequalities follow from the definition of $T$ and the fact $T'<T$, respectively.)
This implies that there exists some $i\in\{1,\ldots,l\}$ for which 
$$
\gamma(\Theta_i) \leq
\frac{\gamma}{2}<\gamma(\Theta_{i-1}).$$ We apply Lemma \ref{lem:cutwhendrops} and obtain that $\cut(\Theta_i)\geq \gamma(\Theta_{i-1})> {\gamma}/{2}$.
Thus, the bottleneck sequence starts with $\cut(\Theta_0)=\cut(I_{T'})<(\gamma/4)+\Delta$ 
(cf.~Eq.~\eqref{eq:cutt}) and eventually {its cut} rises to a value  above $\gamma/2$. From part (ii) of Lemma \ref{prop:bottleneck}, {$\cut(\Theta_i)$} can  increase only when there is a recovery. From part (iii) 
of Lemma \ref{prop:bottleneck}, $\cut(\Theta_i)$ can increase by at most $\Delta$ at each recovery. Thus, in order to obtain an increase from $(\gamma/4)+\Delta$ to $\gamma/2$, we must have had at least $(\gamma/4\Delta)-1$ recoveries in the process $I_t$ between times $T'$ and $T$.

A schematic summary of the two cases introduced in Section \ref{s:l1} is provided in Figure \ref{fig:cases}. 
\begin{figure}
\centering
\includegraphics[width=.8\textwidth]{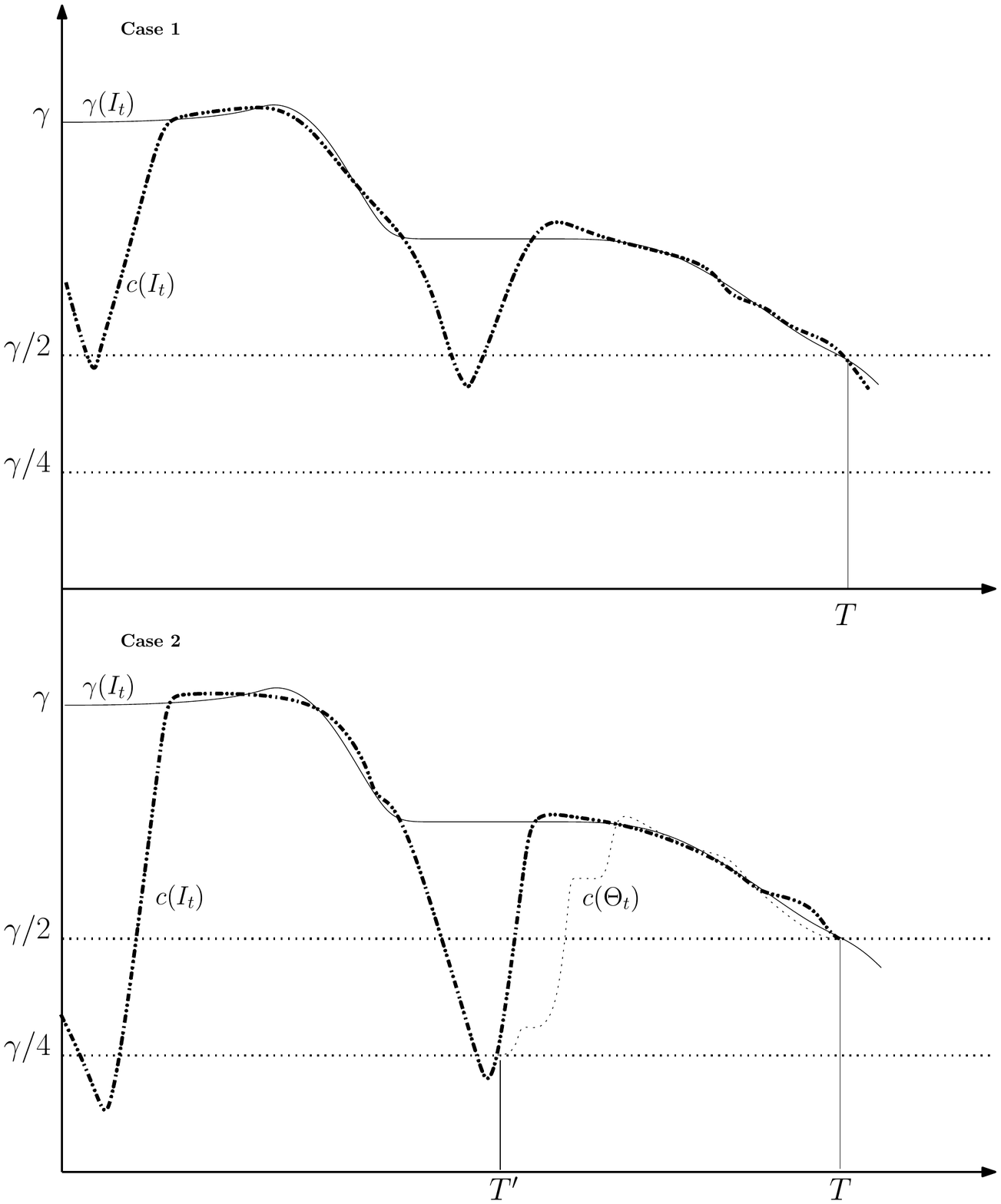}
\caption{\textbf{Case 1:} In the first case,  $\cut(I_t)$ remains at least  $\gamma/4$ throughout the interval $[0,T]$. Moreover, since the resistance drops from $\gamma$ to $\gamma/2$, at least $\gamma/2$ recoveries must occur. \textbf{Case 2:} In the second case, $\cut(I_t)$ drops below $\gamma/4$. The last time that it does so (time $T'$), the resistance is above $\gamma/2$ and needs to drop to a value below $\gamma/2$. Therefore, $\cut(I_t)$ needs to grow above (roughly) $\gamma/2$. In principle,  this increase may happen through infections and not only through recoveries. This is why we define the auxiliary process $\Theta_i$, whose cut also needs to increase to $\gamma/2$ but can only increase through recoveries, implying that at least (roughly) $\gamma/4\Delta$ recoveries occur.
}	 \label{fig:cases}
\end{figure}

\section{Proof of Lemma \ref{biglemma2}.}\label{s:l2}
Lemma \ref{biglemma2} is a fairly routine ``large deviations'' result. It is useful to provide some intuition by considering the special case in which the times $t'$ and $t''$ are fixed (not random), 
{and $\cut(I_t)= \gamma/4$ throughout the interval $[t',t'']$} 
(as opposed to $\cut(I_t)\geq \gamma/4$). 
In this case, we have a Poisson process (recoveries) with rate $r$ and an independent Poisson process (infections) with rate $\gamma/4$; their ratio is $4r/\gamma$. 
For properties (i) and (ii) in Lemma {\ref{biglemma1}} to hold, the empirical ratio of observed recoveries to infections must be at least 
$$\frac{b}{n+b}=\frac{(\gamma/4\Delta) -1}{n+(\gamma/4\Delta)-1},$$ 
{where $b$ is as defined in Lemma \ref{biglemma1}}. When $r$ is small compared to $\gamma/4$, which is the case if we choose $\crr$ small enough, we have an empirical ratio of recoveries to infections which is above  the theoretical ratio by a constant factor. Large deviations theory implies that this event has exponentially small probability. We then argue that within the time horizon of interest, $[0,ce^{cn}]$, there are only $O(ne^{cn})$ intervals that need to be considered. By choosing $c$ small enough and using the union bound, the overall probability that there exist $t'$ and $t''$ with the desired properties can be made small.

The proof for the general case runs along the same lines but involves a coupling argument to show that when $\cut(I_t)$ can exceed $\gamma/4\Delta$, then the event of interest (relatively few infections or, equivalently, too many recoveries) is even less likely to occur. 

\subsection{Decomposing the event of interest.}
Let $c$ be a small enough constant --- how small it needs to be will be seen at the end of the proof. Let $t^*=ce^{cn}$, which is the time horizon of interest in Theorem \ref{th:main}. Recall our definition of the event $B_c$ in Section \ref{s:result}: event $B_c$ occurs if and only if there exists a time interval $[t',t'']$ with $t''\leq ce^{cn}=t^*$, with exactly $b=(\gamma/4\Delta)-1$ recoveries, with at most $n+b$ infections, and during which $\cut(I_t)\geq \gamma/4$. 

Our first step is to show that only a finite number of intervals $[t',t'']$ need to be considered. The recovery process behaves as a Poisson process with rate $r$, as long as the absorbing state has not been entered. To simplify the presentation, let us redefine the process, so that recoveries take place forever, according to a Poisson process. Any recovery that occurs after the extinction time $\tau$ is ``dummy'' and has no effect on the process $\{I_t\}$.

For $i\geq 1$, let $t_i$, be the time of the $i$th  recovery (actual or dummy). 
We consider
the time interval $[t_i,t_{i+b-1}]$, which  is the interval until $b-1$ new recoveries are observed, after the time $t_i$ of the $i$th recovery.

{For $i\geq 1$, we define $B_i$ as the event that} 
throughout  the interval $[t_i,t_{i+b-1}]$ we have $\cut(I_t)\geq \gamma/4$ and at most $n+b$ infections.

\begin{lemma}\label{l:bi}
$\displaystyle{B_c\subseteq \bigcup_{i=1}^{\infty} B_i}$.
\end{lemma}
\proof{Proof:}
Consider a sample path that belongs to $B_c$, so that
there exists an interval $[t',t'']$ with the properties in the definition of $B_c$. In particular, there exists some $i\geq 1$ such that the interval $[t',t'']$ contains the times $t_i,\ldots,t_{i+b-1}$, i.e.,
$$t'\leq t_i\leq t_{i+b-1}\leq t'';$$ 
furthermore, $\cut(I_t)\geq \gamma/4$ during that interval, and we have at most $n+b$ infections.
But in that case, the interval $[t_i,t_{i+b-1}]$ has all of the properties that are required for event $B_i$ to hold.
\Halmos
\endproof

Let $K$ be the total number of  recoveries (real or dummy) during the time interval $[0,t^*]$. Using Lemma \ref{l:bi} and the union bound, we obtain
\begin{equation}\label{eq:b}
\Pr(B_c)\leq \sum_{i=1}^{4rt^*} \Pr(B_i) + \Pr(K>4rt^*)
\leq \sum_{i=1}^{4rt^*} \Pr(B_i) + \frac{1}{4},
\end{equation}
where the last inequality is obtained from the fact that
$K$ is a (Poisson) random variable with mean $rt^*$, and   the Markov inequality.

It remains to bound the sum of the $\Pr(B_i)$. Since $t^*$ grows exponentially with $n$, we are looking for an exponentially small upper bound on each $B_i$. This is the subject of the next subsection.

\subsection{Bounding $\Pr(B_i)$.}\label{s:bound-B}
The main obstacle in characterizing $\Pr(B_i)$ is that the infection process has a time-varying rate. We will handle this issue through a coupling with a Poisson process that has a constant rate.

For $t\geq t_i$, let $M_i(t)$  be the number of infections   during the interval $[t_i, t]$, Let also
\[
C_i(t)=\big \{\cut(I_t) \geq \gamma/4, \, \forall t \in [t_i, t] \big \},
\]
which is the event that $\cut(I_t)$ remains ``large'' during the interval $[t_i,t]$.
Then, the event $B_i$ can be expressed as 
\[
B_i=\big\{M_i(t_{i+b-1})\leq n+b\big\}  \cap  C_i(t_{i+b-1}).
\]

For the remainder of the proof, we assume that $c_r$ is chosen (based only on $\cg$ and $\Delta$, as in the statement of the theorem) so that
\begin{equation}\label{eq:crr}
c_r < \frac{c_{\gamma}^2}{40\Delta}
\end{equation}
By rearranging terms, it is then seen that we can  fix a constant $\bt$ that again depends only on  $\cg$ and $\Delta$, which satisfies
\begin{equation}\label{eq:tb}
c_r\bt<\frac{c_{\gamma}}{5\Delta}\qquad \mbox{and}\qquad \frac{c_{\gamma}}{4} \bt > 2
\end{equation}
For some interpretation and an outline of the rest of the argument,  $\bt$ is chosen so that, with high probability, the interval $[t_i,t_i+\bt]$ has fewer than $b-1$ recoveries, but more than $n+b$ infections if the cut remains ``large.'' As will be seen, this property of $\bt$ implies that, with high probability, the event $B_i$ does not occur.

We  define the event $\barB_i$ by 
\[
\barB_i= \big\{t_{i+b-1} < t_i+\bar{t}\big\} \cup \Big( \big\{M_i( t_i+\bar{t})\leq  n+b \big\} \cap \cut(t_i+\bt)\Big). 
\]
We will now show that $B_i \subseteq \barB_i$.
Consider a sample path in $B_i$. If that sample path also satisfies 
$t_{i+b-1} < t_i+\bar{t}$, then it is also an element of $\barB_i$.
Suppose now that the sample path satisfies $t_{i+b-1} \geq t_i+\bar{t}$.
Using the monotonicity of the counting process $M_i(\cdot)$, we obtain
$M_i( t_i+\bar{t}) \leq M_i(t_{i+b-1}) \leq n+b$, where the last inequality holds because the sample path belongs to $B_i$. Furthermore, since the sample path belongs to $B_i$, it must belong to $C_i(t_{i+b-1})$, which implies that it must also belong to $C_i(t_i+\bt)$. Thus, the sample path belongs to $\big\{M_i( t_i+\bar{t})\leq  n+b \big\} \cap \cut(t_i+\bt)$, and is therefore an element of $\barB_i$. This concludes the proof that $B_i \subseteq \barB_i$. It then follows, using the union bound, that
\begin{equation}
\Pr(B_i) \leq \Pr(\barB_i)\leq  \Pr(t_{i+b-1} < t_i+\bar{t}) + \Pr\Big( \big\{M_i(t_i+\bar{t}) \leq  n+b\big\}\cap \cut(t_i+\bar{t})\Big). \label{eq:bounds}
\end{equation}

Our next step is to derive 
an upper bound for each of the two terms on the right-hand side of Eq.\ \eqref{eq:bounds}, in terms of the Poisson distribution. For the first term, this is simple. The event $\{ t_{i+b-1}< t_i+ \bar{t}\}$ is the event that starting from time $t_i$, at least $b-1$ recoveries occur within $\bt$ time units. Since the recovery process is Poisson with rate $r$, 
we have
\begin{equation}\label{eq:r}
\Pr(t_{i+b-1}< t_i+ \bar{t}) = \Pr\big(R> b-1\big), 
\end{equation}
where $R$ is a Poisson random variable with mean $r\bt$.

To study the second term,  {we use}
$\mathbbm{1}_C$  {to denote}  the indicator function of the event 
$C_i(t_i+\bt)$.
For those sample paths that belong to $C_i(t_i + \bar{t})$, and during the interval $[t_i,t_i+\bt]$, the counting process $M_i(\cdot)$ maintains a rate that is larger than or equal to $\gamma/4$. Thus, on that time interval, $M_i(\cdot)$ can be coupled with a Poisson process $\overline{M}(\cdot)$ with rate equal to $\gamma/4$, in a way that guarantees that
$$M_i(t_i+\bar{t}) \mathbbm{1}_{{C}} \geq
{\overline M}_i(t_i+\bar{t}) \mathbbm{1}_{{C}},$$
for every sample path. Using this dominance relation, we  obtain
\begin{align}
\Pr\Big( \big\{M_i(t_i+\bar{t}) \leq  n+b\big\}\cap \cut(t_i+\bar{t})\Big)
&=\Pr\Big( \big\{M_i(t_i+\bar{t})\mathbbm{1}_{{C}} \leq  n+b\big\}\cap \cut(t_i+\bar{t})\Big) \nonumber\\
&\leq
\Pr\Big( \big\{{\overline M}_i(t_i+\bar{t})\mathbbm{1}_{{C}} \leq  n+b\big\}\cap \cut(t_i+\bar{t})\Big)\nonumber\\
&=
\Pr\Big( \big\{{\overline M}_i(t_i+\bar{t}) \leq  n+b\big\}\cap \cut(t_i+\bar{t})\Big)\nonumber\\
&
\leq \Pr\big(\overline{M}_i(t_i+\bar{t})\leq  n+b \big)\nonumber\\
&= \Pr(\overline{M} \leq  n+b ),
\label{eq:m}
\end{align} 
where $\overline{M}$ is a Poisson random variable with mean $\gamma \bt/4$.

We are now ready to apply large deviations results for Poisson random variables. Note that a Poisson random variable with mean $\lambda n$ can be viewed as a sum of $n$ independent Poisson random variables with mean $\lambda$, and therefore, by the Chernoff bound, the probability of deviating from the mean by a constant factor falls exponentially with $n$. We record this fact in the lemma that follows, which just asserts the fact that we have a positive large deviations exponent.

\begin{lemma}\label{l:ch}
There exists a function $\epsilon(\lambda,\lambda')$, defined for positive $\lambda$ and $\lambda'$, and which is positive whenever $\lambda\neq \lambda'$, with the following properties.
\begin{itemize}
\item[(i)] Let $X$ be a Poisson random variable with mean bounded above by $\lambda n$. If $\lambda'>\lambda$, then
$$\Pr(X\geq \lambda' n)\leq e^{-\epsilon(\lambda,\lambda') n},
\qquad {\forall\ n.}$$
\item[(ii)] Let $X$ be a Poisson random variable with mean bounded below by $\lambda n$. If $\lambda'<\lambda$, then
$$\Pr(X\leq \lambda' n)\leq e^{-\epsilon(\lambda,\lambda') n},
\qquad {\forall\ n.}$$
\end{itemize}
\end{lemma}

The random variable $R$ in Eq.\ \eqref{eq:r} is Poisson with mean $r\bt\leq c_r \bt n$. 
Note that, for large enough $n$, we have $b-1=(\gamma/4\Delta)-2\geq(\gamma/5\Delta) \geq (c_{\gamma}/5\Delta)n$, {where the last inequality follows from the fact that $\gamma\geq c_\gamma n$}. .
We  apply  Lemma \ref{l:ch}(i), with $\lambda=c_r\bt$ and $\lambda'=c_{\gamma}/5\Delta$:
$$\Pr\big( R >b{-1}\big) \leq 
\Pr\big(  R > (c_{\gamma}/5\Delta)n \big)
\leq e^{-\epsilon_1 n}.$$
Because of our assumptions on $c_r$ and $\bt$ (cf.\ Eq.\ \eqref{eq:tb}),
we have $\lambda'>\lambda'$, and $\epsilon_1$ is a positive number determined by $c_r$, $c_{\gamma}$, and $\Delta$.

Similarly, the random variable $\oM$ in Eq.\ \eqref{eq:m} is Poisson with mean $\gamma \bt /4\geq (c_{\gamma} \bt /4)n$. 
For any graph, $\gamma$ is bounded above by $n\Delta$, and this implies that $b=(\gamma/4\Delta)-1 \leq n$.
We  apply  Lemma \ref{l:ch}(ii), with $\lambda=c_{\gamma} \bt /4$ 
and $\lambda'=2$:
$$\Pr\big(\oM \leq n+b \big) \leq \Pr\big(\oM \leq 2n \big)
\leq e^{-\epsilon_2 n}.$$
Because of our assumptions on $c_{\gamma}$ and $\bt$ (cf.\ Eq.\  \eqref{eq:tb}),
we have $\lambda'<\lambda$, and $\epsilon_2$ is a positive number determined by $c_{\gamma}$.

We have therefore established that each of the two terms on the right-hand side of Eq.\ \eqref{eq:bounds} is bounded above by an exponentially decaying term. By letting $\epsilon=\min\{\epsilon_1,\epsilon_2\}>0$, we obtain that
\begin{equation}\label{eq:bibd}
\Pr(B_i) \leq 2 e^{-\epsilon n}.
\end{equation}

\subsection{Completing the proof of Lemma \ref{biglemma2}.} For the given $\cg$ and $\Delta$, we choose  a suitably small $\crr$ as in 
Eq.\ \eqref{eq:crr}. This allows us to set $\bt$ as in Eq.\ \eqref{eq:tb}, leading to a positive $\epsilon$ in Eq.\ \eqref{eq:bibd}.
We then use Eq.\ \eqref{eq:bibd} to bound the terms $\Pr(B_i)$ in the inequality
\eqref{eq:b}, and also make use of the facts that
$t^*=ce^{cn}$ and $4rt^* \leq 4 \crr n c e^{cn}$,
to obtain
$$\Pr(B_c)\leq 4\crr  n c e^{cn}  2 e^{-\epsilon n} + \frac{1}{4}
\leq \frac{1}{2},$$
provided that $c$ is small enough (it just needs to be chosen a little smaller than $\epsilon$) and $n$ is large enough.
This concludes the proof of Lemma \ref{biglemma2}.

\section{Conclusions.}\label{s:con}
We have considered the 
control of an epidemic (contagion process) given a limited curing budget, and provided an exponential lower bound on the expected time to extinction, 
{for bounded degree graphs.}
{For the interesting (and least favorable) case where all nodes are initially infected, our assumption was}
that the CutWidth of the graph scales linearly with the number of nodes, and 
{that the curing budget is bounded above by a small enough multiple of the number of nodes.}

This result complements the results in \cite{DOT14}, which show that when the ratio of the curing budget to the CutWidth is large enough, then the expected time to extinction is sublinear in the number of nodes. These results, taken together, show that for graphs with a large CutWidth, the ratio of the curing resources to the CutWidth is the key factor that distinguishes between slow and fast extinction. 
Our proof was based on a generalization of the CutWidth, the ``resistance,'' which captures the difficulty of extinguishing an epidemic, starting from an arbitrary set of infected nodes. 

It remains an open problem to develop lower bounds for more general bounded-degree graphs, whose CutWidth scales sublinearly with the number of nodes. In some cases, this is easy. For example, for a square mesh with $n$ nodes, the CutWidth is of order $O(\sqrt{n})$. Using the fact that any subset of the mesh with $\Theta(n)$ nodes has a cut of size $\Omega(\sqrt{n})$, one can show that a curing budget that scales at least as fast as $\sqrt{n}$ is necessary for fast exticntion. 
The same argument applies whenever we deal with families of graphs that satisfy 
suitable isoperimetric inequalities. We conjecture that a similar result is always  true: that is, unless the curing budget scales in proportion with the CutWidth, the expected time to extinction will be exponential. However, some new tools may have to be developed.

The proof of Theorem \ref{th:main}, and in particular Eq.\ \eqref{eq:crr}, shows that the exponential lower bound holds when $c_r$ is smaller than a constant multiple of $c_{\gamma}^2$. We conjecture that a similar lower bound can be established {under the assumption that} $c_r$ is smaller than a constant multiple of $c_{\gamma}$. If this is true, the deciding factor will be the ratio between the resistance and the recovery rate {in a very concrete  sense.} However, the proof of this conjecture, if true,  will require a much more refined argument.

Finally, the problem of controlling contagion processes on networks gives rise to a broader family of  interesting research directions, such as control under partial information on the state of each node, combining inference and control etc. 

% Acknowledgments here
\section*{Acknowledgments.} This research was partially supported by NSF under grant CMMI-1234062, by the Draper Laboratories and by ARO under grant W911NF-12-1-0509.

% Enter the text of acknowledgments here

% References here (outcomment the appropriate case) 

% CASE 1: BiBTeX used to constantly update the references 
%   (while the paper is being written).
%\bibliographystyle{ormsv080} % outcomment this and next line in Case 1
%\bibliography{<your bib file(s)>} % if more than one, comma separated

% CASE 2: BiBTeX used to generate mypaper.bbl (to be further fine tuned)
%\input{mypaper.bbl} % outcomment this line in Case 2

 \bibliography{epid}
 \bibliographystyle{abbrv}

\begin{APPENDICES}

\section{Proof of Lemma \ref{lem:infectiondelta}.}
\label{ap:lemma}

{Recall that $\Omega(A)$ stands for the set of all 
($A$-$\emptyset)$-crusades. Let also
$\Omega^A$ be the set of all such crusades  that achieve the minimum in the definition of the resistance, i.e., 
\[
\Omega^A=\{ \omega \in \Omega(A): z(\omega)=\gamma(A) \}.
\]}
\begin{enumerate}[(i)]
\item  Suppose that $A\subseteq B$. Let $\omega^B=(\omega^B_0, \ldots, \omega^B_k)\in \Omega^B$. Consider the sequence $\hat{\omega}=(\homega_0, \ldots, \homega_k)$ of bags with $\hat{\omega}_0 = A$, and $\hat{\omega}_i =\omega^B_i$, for $i = 1, \ldots, k$.  We claim that $\hat{\omega}$ is a crusade $\hat{\omega}\in \Omega(A)$. Indeed,
 \begin{enumerate}
 \item $\hat{\omega}_0=A$;
 \item $\hat{\omega}_k=\omega^B_k=\emptyset$;
 \item $|\hat{\omega}_0\setminus \hat{\omega}_1|=|A\setminus\hat{\omega}_1|\leq|B \setminus{\omega^B_1}|=|{\omega^B_0} \setminus{\omega^B_1}|\leq 1$, where the first inequality follows from  $A \subseteq B$ and $\hat{\omega}_1= {\omega^B_1}$. Moreover, for  $i = {0, \ldots, k-1}$, we have  $|\hat{\omega}_i
 \setminus\hat{\omega}_{i+1}|=|{\omega^B_i}
  \setminus 
 {\omega^B_{i+1}}|\leq 1$.
\end{enumerate}
Clearly,
\[
z(\hat{\omega})=\max_{ 1\leq i \leq k}    \{\cut(\hat{\omega}_i )\} =\max_{ 1\leq i \leq k}\{\cut(\omega^B_i )\} = \gamma(B).
\]
\noindent Using {the} definition of $\gamma(A)$, and the fact that $\hat\omega\in\Omega(A)$, we conclude that
\[
\gamma(A)= \min_{\omega \in \Omega(A)} z(\omega) \leq z(\homega)= \gamma(B).
\]

\item   If $|A\triangle B|=m$, we can go from bag $A$ to bag $B$ in a sequence of $m$ steps, where at each step, we add or remove a single node. It thus suffices to show that each one of these steps can change the resistance by at most $\Delta$. Accordingly, we only need to conside the case where $B=A+v$, for some $v\notin A$.
 
Let $\omega^A=(\omega^A_0, \ldots, \omega^A_k)\in \Omega^A$.  Consider the sequence $\hat{\omega} = (\homega_0, \ldots, \homega_{k+1})$ of bags  {with}  $\hat{\omega}_i ={\omega^A_i} +v$, for  $i= 0, \ldots, {k }$, and $\hat{\omega}_{k+1}=\emptyset$. Clearly, $\hat{\omega}$ is a crusade in $\Omega(B)$ and, therefore, 
 \[
\gamma(B) \leq z(\homega) 
{=\max_{1\leq i\leq k}\{\cut(\omega_i^A +v)\}}
\leq \max_{ 1\leq i \leq k}    \{\cut(\omega^A_i )\} + \Delta= \gamma(A) + \Delta,
 \] 
 where the second inequality follows because
the addition of one node can  change the cut  by at most $\Delta$ (Lemma  \ref{prop:cutproperties}(i)).
   \end{enumerate}
\end{APPENDICES}
\end{document}